\documentclass[onecolumn]{article}
\usepackage{amssymb}

\input{tcilatex}

\begin{document}

\title{Short-time decoherence of Josephson charge qubit nonlinearly coupling
with its environment}
\author{Xian-Ting Liang\thanks{%
Corresponding author. E-mail address: xtliang@ustc.edu (X.T.L).}, Yong-Jian
Xiong \\
Department of Physics and Institute of Modern Physics,\\
Ningbo University, Ningbo 315211, China}
\maketitle

\begin{abstract}
At first, we generally investigate the short-time decoherence of a qubit
nonlinearly coupling with a bath. The measure of the decoherence is chosen
as the maximum norm of the deviation density operator. Then we concretely
investigate the Josephson charge qubit (JCQ) model. It is shown that at the
temperature $T\sim 30mK,$ the loss of fidelity (due to decoherence) of the
JCQ is bigger than the DiVincenzo low decoherence criterion. The decoherence
will decrease with the decrease of the experimental temperature. When the
temperature decreases to $T\sim 0.3mK$ the DiVincenzo low decoherence
criterion can be satisfied.

PACS numbers: 03.65.Ta, 03.65.Yz, 85.25.Cp

Keywords: Short-time decoherence, Josephson junction, quantum computation
\end{abstract}

\section{Introduction}

David DiVincenzo put forward a low decoherence criterion for the candidates
of quantum computing hardware to be satisfied \cite{DiVincenzo}. An
approximate benchmark of the criterion is a fidelity loss no more than $\sim
10^{-4}$ per elementary quantum gate operation. In general, the loss of the
fidelity results from quantum leakage as well as decoherence. The Josephson
junction is considered to be a promising physical realization of the qubit.
It has been shown that the quantum leakage of Josephson charge qubit (JCQ)
is not serious \cite{PRL83-5385-1999}. So it is very interesting to
investigate the decoherence of the JCQ. To perform the quantitative study of
the decoherence for a qubit, in general one needs to solve the quantum
dynamical problem of the qubit coupling with its environment. However, the
solutions of the coupling system can not be obtained easily, usually a kind
of approximation must be appealed. The Markovian approximation \cite{Markov}
is the most familiar one. This approximation has been used to study the
decoherence of qubits in last years \cite{chemphys268-73-2001} \cite%
{PRB68-060502-2003}. But it is not a suitable tool for researching the
short-time decoherence of the qubits \cite{Privman}. However, the short-time
processes is very important because almost all of the gate operations of the
quantum devices are short time. Fortunately, a short-time approximation
scheme for the reduced density matrix has been developed recently \cite%
{Privman}. Another core of the issue to investigate the decoherence of qubit
is to make certain what the model of the environment and the coupling form
of the qubit with its environment are. In general, according to the
characteristic of the environment we set it a bosonic (or fermi) bath. The
coupling of the qubit-bath can be supposed linear because the interaction
between the qubit and the modes of the bath is very weak. Under these
suppositions (linear coupling and short-time approximation), V. Privman 
\emph{et al. }\cite{Privman} \cite{Privman-PRA} \cite{Fedichkin et al}
investigated the decoherence of many kinds of open systems and we also
investigated a concrete JCQ model in Ref.\cite{Liang01}. However, the
nonlinearly coupling forms of qubit-bath certainly exists \cite%
{PhysicaE18-2003-29}. In this Letter, we firstly introduce a model that JCQ
nonlinearly couples with its bath and then use the short-time approximation
to investigate the decoherence of the JCQ in the qubit-bath model.

\section{Nonlinearly coupling qubit-bath model}

Though the combined effects of the bath modes on the qubit may be large
enough, the coupling of the qubit with each of the modes of the bath is very
weak and the qubit-bath can be considered a linear coupling. However, due to
the development of low temperature technique and fabrication technique of
high quality quantum systems the nonlinearly coupling qubit-bath models can
not be excluded in the devices of quantum. The linearly coupling qubit-bath
model is analogous to the amplitude damping model. In this Letter, we
introduce a nonlinearly coupling qubit-bath model which corresponds to the
phase damping model \cite{Preskill}. We begin our investigation by
considering a general open quantum system%
\begin{equation}
H=H_{S}+H_{B}+H_{I}.  \label{e1}
\end{equation}%
Here, $H_{S}$ is the Hamiltonian of the investigated quantum system; $H_{B}$
is the Hamiltonian of the bath; and $H_{I}$ describes the interaction of the
system and the bath. The bath is traditionally modeled by a large number of
uncoupled harmonic oscillators (with ground energy shifted to zero),%
\begin{equation}
H_{B}=\sum_{j}\omega _{j}b_{j}^{\dagger }b_{j}.  \label{e2}
\end{equation}%
If the interaction is linear, in the rotating-wave approximation the
interaction Hamiltonian is%
\begin{equation}
H_{I}=\Lambda \sum_{j}g_{j}b_{j}^{\dagger }+\Lambda ^{\dagger
}\sum_{j}g_{j}^{\ast }b_{j},  \label{e3}
\end{equation}%
with the interaction constants $g_{j}.$ If the investigated system is a
two-level system then $\Lambda \rightarrow \sigma _{-},$ $\Lambda ^{\dagger
}\rightarrow \sigma _{+}$. The short-time decoherence of this model has been
investigated in Ref.\cite{Privman-PRA}. Suppose the bath be ohmic then%
\begin{equation}
g^{2}\left( \omega \right) D\left( \omega \right) =\eta \omega \exp \left(
\omega /\omega _{c}\right) ,  \label{e4}
\end{equation}%
where $D\left( \omega \right) $ is the density of states of the bath and $%
\omega _{c}$ is the cutoff frequency. For the model of JCQ coupling with the
bath (see Fig.1 of Ref.\cite{chemphys259-1-2003}), the damping coefficient
can be calculated with 
\begin{equation}
\eta =\frac{R}{R_{Q}}\left( \frac{C_{t}}{C_{J}}\right) ^{2},  \label{e5}
\end{equation}%
where $R_{Q}=\left( 2e\right) ^{2}/h$ is the (superconducting) resistance
quantum. When we investigate the short-time decoherence of this model, we
can directly use $g^{2}\left( \omega \right) D\left( \omega \right) $ and do
not need solving $D\left( \omega \right) $ then $g\left( \omega \right) .$

As well known that the interaction Hamiltonian of a oscillator in the phase
damping bath is \cite{Preskill} \cite{ph/0408087}%
\begin{equation}
H_{I}=\sum_{j}\left( g_{j}b_{j}^{\dagger }b_{j}\right) a^{\dagger }a,
\label{e6}
\end{equation}%
which is a nonlinear coupling. If we replace the oscillator with a two-level
system, namely, $a^{\dagger }\rightarrow \sigma _{+},$ $a\rightarrow \sigma
_{-},$ then the total Hamiltonian becomes%
\begin{equation}
H=H_{S}+\sigma _{z}\sum_{j}g_{j}^{\prime }b_{j}^{\dagger
}b_{j}+\sum_{j}\omega _{j}^{\prime }b_{j}^{\dagger }b_{j},  \label{e7}
\end{equation}%
where $g_{j}^{\prime }=g_{j}/2$ and $\omega _{j}^{\prime }=g_{j}/2+\omega
_{j}.$ This Hamiltonian models a kind of nonlinear coupling of the bath to
the exposed qubit. This bath can be considered resulting from the
fluctuations of the Josephson nanocircuits and the surrounding circuits. A
more general model which includes other sources of decoherence, such as
quasiparticle tunneling, fluctuating background charges and flux noise has
been put forward \cite{PhysicaE18-2003-29} \cite{PRL88-047901-2002}. Suppose
the bath is also ohmic then Eq.(\ref{e4}) will be held. But in this case, in
order to investigate the decoherence of the qubit we must find out $g\left(
\omega _{j}\right) $ through solving $D\left( \omega _{j}\right) $. We
suppose that the initial state of the environment state is $\Theta $ which
is the product of the bath modes density matrices%
\begin{equation}
\theta _{j}=\frac{e^{-\beta M_{j}}}{\text{Tr}_{j}\left( e^{-\beta
M_{j}}\right) },  \label{e8}
\end{equation}%
where $M_{j}=\omega _{j}b_{j}^{\dagger }b_{j}.$ Because the degrees of
freedom of the bath are much larger than that of the qubit, in the short
time we can take the bath's states unchanged and consider the bath keeping
in their thermal equilibrium state. We take the bath to be a mix of infinite
harmonic oscillators at temperature $T.$ Thus, the density of state of the
bath reads 
\begin{equation}
D\left( \omega _{j}\right) =\frac{1-\exp \left( -\hbar \omega _{j}\right) }{%
\exp \left( \beta \hbar j\omega _{j}\right) }\approx \hbar \omega _{j},
\label{e9}
\end{equation}%
for $kT\ll \hbar \omega _{j}$. Here $\beta =1/kT$, $k$ is the Boltzmann
constant. By using Eq.(\ref{e4}) we have%
\begin{eqnarray}
g^{\prime }\left( \omega _{j}\right) &=&\frac{1}{2}\sqrt{\frac{\hbar \eta
\omega _{j}}{\sqrt{D\left( \omega _{j}\right) }}}\exp \left( \omega
_{j}/2\omega _{c}\right)  \nonumber \\
&\approx &\frac{1}{2}\sqrt{\eta }\exp \left( \omega _{j}/2\omega _{c}\right)
.  \label{e10}
\end{eqnarray}

\section{Short-time dynamics of the qubit in the nonlinear qubit-bath model}

In this section we will investigate the general expression for the time
evolution operator of the qubit in the nonlinear qubit-bath model within the
short time. The definition of the short-time concept has been stated in Ref. 
\cite{Privman} \cite{Privman-PRA} \cite{Fedichkin et al}. The evolution
operator is%
\begin{equation}
U=e^{-iH\tau /\hslash }=e^{-i\left( H_{s}+H_{I}+H_{B}\right) \tau /\hslash }.
\label{e11}
\end{equation}%
In the following we set $t=\tau /\hslash .$ Due to non-conservation of $%
H_{s} $ in this system, the evolution operator cannot be in a general way
expressed as $e^{-iH_{s}t}e^{-i\left( H_{I}+H_{B}\right) t}.$ But in the
sort-time approximation, the operator can be approximately expressed as \cite%
{split-operator01} \cite{split-operator02}%
\begin{equation}
U=e^{-iH_{s}t/2}e^{-i\left( H_{I}+H_{B}\right) t}e^{-iH_{s}t/2}+o(t^{3}).
\label{e12}
\end{equation}%
It has been proved that the expression is accurate enough for the time being
short to the characteristic time. So the density matrix elements of the
reduced density matrix $\rho \left( t\right) $ in the basis of operator $%
H_{s}$ can be expressed as%
\begin{equation}
\rho _{mn}=\text{Tr}_{B}\left\langle \varphi _{m}\right\vert
e^{-iH_{s}t/2}e^{-i\left( H_{I}+H_{B}\right) t}e^{-iH_{s}t/2}R\left(
0\right) e^{iH_{s}t/2}e^{i\left( H_{I}+H_{B}\right)
t}e^{iH_{s}t/2}\left\vert \varphi _{n}\right\rangle .  \label{e13}
\end{equation}%
Here, we suppose that the initial state of the system is $R\left( 0\right)
=\rho \left( 0\right) \otimes \Theta ,$ where $\rho \left( 0\right) $ is the
initial state of the qubit and $\Theta =\tprod \theta _{j}$. By using the
completeness relation of the eigenstates of $H_{s},$ $\tsum \left\vert \cdot
\right\rangle \left\langle \cdot \right\vert =1$, we have 
\begin{equation}
e^{\pm iH_{s}t/2}=\tsum_{j=0,1}e^{\pm it\lambda _{j}}\left\vert \varphi
_{j}\right\rangle \left\langle \varphi _{j}\right\vert ,  \label{e14}
\end{equation}%
where $\lambda _{j}$ $\left( j=0,1\right) $ and $\left\vert \varphi
_{j}\right\rangle $ are eigenvalues and eigenstates of Hamiltonian $H_{s}$.
Similarly, we have 
\begin{equation}
e^{\pm i\left( H_{I}+H_{B}\right) t}=\tsum_{j=0,1}\exp \left\{ \sum_{j}\left[
\pm i\left( g_{j}^{\prime }\chi _{j}+\omega _{j}^{\prime }\right)
b_{j}^{\dagger }b_{j}t\right] \right\} \left\vert \psi _{j}\right\rangle
\left\langle \psi _{j}\right\vert ,  \label{e15}
\end{equation}%
where $\chi _{j}$ $\left( j=0,1\right) $ and $\left\vert \psi
_{j}\right\rangle $ $\left( j=0,1\right) ,$ are the eigenvalues and
eigenstates of operator $\sigma _{z}.$ So we have%
\begin{eqnarray}
\rho _{mn}\left( t\right) &=&\text{Tr}_{B}\left\{ \left\langle \varphi
_{m}\right\vert \tsum_{\alpha =0,1}e^{-it\lambda _{\alpha }}\left\vert
\varphi _{\alpha }\right\rangle \left\langle \varphi _{\alpha }\right\vert
\right.  \nonumber \\
&&\tsum_{\xi =0,1}\exp \left\{ \sum_{j}\left[ -i\left( g_{j}^{\prime }\chi
_{\xi }+\omega _{j}^{\prime }\right) b_{j}^{\dagger }b_{j}t\right] \right\}
\left\vert \psi _{\xi }\right\rangle \left\langle \psi _{\xi }\right\vert 
\nonumber \\
&&\tsum_{\beta =0,1}e^{-it\lambda _{\beta }}\left\vert \varphi _{\beta
}\right\rangle \left\langle \varphi _{\beta }\right\vert
\tsum_{p,q=0,1}\left\vert \varphi _{p}\right\rangle \left\langle \varphi
_{q}\right\vert  \nonumber \\
&&\times \rho _{pq}\left( 0\right) \tprod_{j}\theta _{j}\tsum_{\mu
=0,1}e^{it\lambda _{\mu }}\left\vert \varphi _{\mu }\right\rangle
\left\langle \varphi _{\mu }\right\vert  \nonumber \\
&&\tsum_{\varsigma =0,1}\exp \left\{ \sum_{j}\left[ i\left( g_{j}^{\prime
}\chi _{\varsigma }+\omega _{j}^{\prime }\right) b_{j}^{\dagger }b_{j}t%
\right] \right\} \left\vert \psi _{\varsigma }\right\rangle \left\langle
\psi _{\varsigma }\right\vert  \nonumber \\
&&\left. \tsum_{\nu =0,1}e^{it\lambda _{\nu }}\left\vert \varphi _{\nu
}\right\rangle \left\langle \varphi _{\nu }\right\vert \left. \varphi
_{n}\right\rangle \right\} ,  \label{e16}
\end{eqnarray}%
namely,%
\begin{eqnarray}
\rho _{mn}\left( t\right) &=&\tsum_{\alpha ,\beta ,\xi ,\varsigma ,p,q,\mu
,\nu =0,1}e^{it\left( \lambda _{\mu }+\lambda _{\nu }-\lambda _{\alpha
}-\lambda _{\beta }\right) }\left\langle \varphi _{m}\right\vert \left.
\varphi _{\alpha }\right\rangle  \nonumber \\
&&\left\langle \varphi _{\alpha }\right\vert \left. \psi _{\xi
}\right\rangle \left\langle \psi _{\xi }\right\vert \left. \varphi _{\beta
}\right\rangle \left\langle \varphi _{\beta }\right\vert \left. \varphi
_{p}\right\rangle \left\langle \varphi _{q}\right\vert \left. \varphi _{\mu
}\right\rangle \left\langle \varphi _{\mu }\right\vert \left. \psi
_{\varsigma }\right\rangle \left\langle \psi _{\varsigma }\right\vert \left.
\varphi _{\nu }\right\rangle  \nonumber \\
&&\left\langle \varphi _{\nu }\right\vert \left. \varphi _{n}\right\rangle
\rho _{pq}\left( 0\right) \text{Tr}_{B}\left[ \exp \left\{ \sum_{j}\left[
-i\left( g_{j}^{\prime }\chi _{\xi }+\omega _{j}^{\prime }\right)
b_{j}^{\dagger }b_{j}t\right] \right\} \right.  \nonumber \\
&&\left. \tprod_{j}\theta _{j}\exp \left\{ \sum_{j}\left[ i\left(
g_{j}^{\prime }\chi _{\varsigma }+\omega _{j}^{\prime }\right)
b_{j}^{\dagger }b_{j}t\right] \right\} \right] .  \label{e17}
\end{eqnarray}%
Taking the $j$ term in Tr$_{B}\left[ \text{...}\right] ,$ and denoting it by
Tr$_{B}\left[ \text{...}\right] _{j}$, we have%
\begin{eqnarray}
Tr_{B}\left[ \text{...}\right] _{j} &=&\text{Tr}_{B}\left\{ \exp \left[
-i\left( g_{j}^{\prime }\chi _{\xi }+\omega _{j}^{\prime }\right)
b_{j}^{\dagger }b_{j}t\right] \theta _{j}\exp \left[ i\left( g_{j}^{\prime
}\chi _{\varsigma }+\omega _{j}^{\prime }\right) b_{j}^{\dagger }b_{j}t%
\right] \right\}  \nonumber \\
&=&\text{Tr}_{B}\left\{ \exp \left[ -i\left( g_{j}^{\prime }\chi _{\xi
}+\omega _{j}^{\prime }\right) b_{j}^{\dagger }b_{j}t\right] \right. \exp
\left( -\beta \omega _{j}b_{j}^{\dagger }b_{j}\right)  \nonumber \\
&&\left[ \text{Tr}\exp \left( -\beta \omega _{j}b_{j}^{\dagger }b_{j}\right) %
\right] ^{-1}\left. \exp \left[ i\left( g_{j}^{\prime }\chi _{\varsigma
}+\omega _{j}^{\prime }\right) b_{j}^{\dagger }b_{j}t\right] \right\} .
\label{e18}
\end{eqnarray}%
Because%
\begin{equation}
\exp \left[ xb_{j}^{\dagger }b_{j}t\right] =:\exp \left[ \left(
e^{x}-1\right) b_{j}^{\dagger }b_{j}\right] :,  \label{e19}
\end{equation}%
where $:\ :$ denotes the normally ordering form. By using the relationship
of the completeness of the coherent states%
\begin{eqnarray}
\int \frac{d^{2}z}{\pi }\left\vert z\right\rangle \left\langle z\right\vert
&=&1,  \nonumber \\
\int \frac{d^{2}\xi }{\pi }\left\vert \xi \right\rangle \left\langle \xi
\right\vert &=&1,  \nonumber \\
\int \frac{d^{2}\xi }{\pi }\left\vert \varsigma \right\rangle \left\langle
\varsigma \right\vert &=&1,  \label{e20}
\end{eqnarray}%
we have%
\begin{eqnarray}
Tr_{B}\left[ \text{...}\right] _{j} &=&\frac{1}{Z_{j}}\int \frac{d^{2}z}{\pi 
}\left\langle z\right\vert :\exp \left[ \left( e^{-i\left( g_{j}^{\prime
}\chi _{\xi }+\omega _{j}^{\prime }\right) t}-1\right) b_{j}^{\dagger }b_{j}%
\right] :  \nonumber \\
&:&\exp \left[ \left( e^{-\beta \omega _{j}}-1\right) b_{j}^{\dagger }b_{j}%
\right] ::\exp \left[ \left( e^{i\left( g_{j}^{\prime }\chi _{\xi }+\omega
_{j}^{\prime }\right) t}-1\right) b_{j}^{\dagger }b_{j}\right] :\left\vert
z\right\rangle  \nonumber \\
&=&\frac{1}{Z_{j}}\int \frac{d^{2}z}{\pi }\int \frac{d^{2}\xi }{\pi }\int 
\frac{d^{2}\varsigma }{\pi }\exp \left( -\left\vert z\right\vert
^{2}-\left\vert \xi \right\vert ^{2}-\left\vert \varsigma \right\vert
^{2}\right)  \nonumber \\
&&\exp \left[ e^{-i\left( g_{j}^{\prime }\chi _{\xi }+\omega _{j}^{\prime
}\right) t}z^{\ast }\xi +e^{-\beta \omega _{j}}\xi ^{\ast }\varsigma
+e^{i\left( g_{j}^{\prime }\chi _{\xi }+\omega _{j}^{\prime }\right)
t}\varsigma ^{\ast }z\right] ,  \label{e21}
\end{eqnarray}%
where $Z_{j}=\left( 1-e^{-\beta \omega _{j}}\right) ^{-1}.$ By using the
formula%
\begin{eqnarray}
&&\int \frac{d^{2}z}{\pi }\exp \left( a\left\vert z\right\vert
^{2}+bz+cz^{\ast }+fz^{2}+gz^{\ast 2}\right)  \nonumber \\
&=&\frac{1}{\sqrt{a^{2}-4fg}}\exp \left[ \frac{-abc-b^{2}g-c^{2}f}{\sqrt{%
a^{2}-4fg}}\right] ,  \label{e22}
\end{eqnarray}%
we can obtain the result of $\rho _{mn}\left( t\right) $ as%
\begin{eqnarray}
\rho _{mn}\left( t\right) &=&\tsum_{\alpha ,\beta ,\xi ,\varsigma ,p,q,\mu
,\nu =0,1}e^{it\left( \lambda _{\mu }+\lambda _{\nu }-\lambda _{\alpha
}-\lambda _{\beta }\right) }  \nonumber \\
&&\left\langle \varphi _{m}\right\vert \left. \varphi _{\alpha
}\right\rangle \left\langle \varphi _{\alpha }\right\vert \left. \psi _{\xi
}\right\rangle \left\langle \psi _{\xi }\right\vert \left. \varphi _{\beta
}\right\rangle \left\langle \varphi _{\beta }\right\vert \left. \varphi
_{p}\right\rangle \left\langle \varphi _{q}\right\vert \left. \varphi _{\mu
}\right\rangle \left\langle \varphi _{\mu }\right\vert \left. \psi
_{\varsigma }\right\rangle  \nonumber \\
&&\left\langle \psi _{\varsigma }\right\vert \left. \varphi _{\nu
}\right\rangle \left\langle \varphi _{\nu }\right\vert \left. \varphi
_{n}\right\rangle \rho _{pq}\left( 0\right) \dprod\limits_{j=1}^{N}\frac{%
1-e^{-\beta \omega _{j}}}{1-e^{ig_{j}^{\prime }\left( \chi _{\zeta }-\chi
_{\xi }\right) t}e^{-\beta \omega _{j}}}.  \label{e23}
\end{eqnarray}%
It should be noted that it is not necessary to take $N\ $too large in the
following numerical simulations because the $\omega _{j}$ increase with $j$
and the $e^{-\beta \omega _{j}}$ decrease with $\omega _{j}$ \emph{%
exponentially}. By using Eq.(\ref{e23}) we can investigate various
characteristics of the nonlinear qubit-bath systems. By using it, in the
following we shall study the short-time decoherence of the JCQ.

\section{Measure of decoherence for a qubit}

In order to investigate the decoherence of the qubits, one must designate
the measure to be used. There are many choices for this purpose. When the
evolution time is very long, the qubit interacting with the large bath will
fall into the thermal equilibrium at temperature $T$. In this case,
Markovian type approximation can be used to quantify the decoherent
processes and it usually yields the exponential decay of the density matrix
elements in the energy basis of the Hamiltonian $H_{s}.$ In this time scale
the measures of entropy and the first entropy can be used to quantify the
decoherence. But the decoherence of the qubit gates cannot be characterized
by this methods because the time of the elementary quantum gate operations
are much shorter than the thermal relaxation time. It has been shown that
the norm $\left\Vert \sigma \right\Vert _{\lambda }$ is useful for
describing the decoherence of the short-time evolutions. In this section, we
will review the measure according to \cite{Privman}. We set $\sigma $ is the
deviation operator defined as%
\begin{equation}
\sigma \left( t\right) =\rho \left( t\right) -\rho ^{i}\left( t\right) ,
\label{e24}
\end{equation}%
where $\rho \left( t\right) $ and $\rho ^{i}\left( t\right) $ are density
matrixes of the \textquotedblleft real\textquotedblright\ evolution (with
interaction) and the \textquotedblleft ideal\textquotedblright\ one (without
interaction) of the investigated system. $\left\Vert \sigma \right\Vert
_{\lambda }$ is defined as 
\begin{equation}
\left\Vert \sigma \right\Vert _{\lambda }=\sup_{\varphi \neq 0}\left( \frac{%
\left\langle \varphi \right\vert \sigma \left\vert \varphi \right\rangle }{%
\left\langle \varphi \right. \left\vert \varphi \right\rangle }\right) ^{%
\frac{1}{2}}.  \label{e25}
\end{equation}%
For a qubit, the norm can be given by%
\begin{equation}
\left\Vert \sigma \right\Vert _{\lambda }=\sqrt{\left\vert \sigma
_{10}\right\vert ^{2}+\left\vert \sigma _{11}\right\vert ^{2}}.  \label{e26}
\end{equation}%
In general, the norm $\left\Vert \sigma \right\Vert _{\lambda }$ increases
with time, reflecting the decoherence of the system. However, it is
oscillating at the system's internal frequency. Thus, the decohering effect
of the bath is better quantified by the maximal operator norm, $D\left(
t\right) $ which is defined as 
\begin{equation}
D\left( t\right) =\sup_{\rho \left( 0\right) }\left( \left\Vert \sigma
\left( t,\rho \left( 0\right) \right) \right\Vert _{\lambda }\right) .
\label{e27}
\end{equation}%
It has been shown that the measure can rightly describe the short-time
decoherence of qubits. In the following we will use this measure to
investigate the short-time decoherence of JCQ in nonlinearly coupling
JCQ-bath model.

\section{Short-time decoherence of the JCQ}

In this section we firstly review the JCQ model. The single JCQ Hamiltonian
is \cite{RMP2001}%
\begin{equation}
H_{R}=E_{ch}\left( n-n_{g}\right) ^{2}-E_{J}\cos \varphi .  \label{e28}
\end{equation}%
Here, $E_{ch}=2e^{2}/\left( C_{g}+C_{J}\right) $ is the charging energy; $%
E_{J}=\Phi _{0}I_{c}/2\pi $ is the Josephson coupling energy, where $I_{c}$
is the critical current of the Josephson junction, $\Phi _{0}=hc/2e$ denotes
the flux quantum; $n_{g}=C_{g}V_{g}/2e$ is the dimensionless gate charge,
where $C_{g}$ is the gate capacitance, $V_{g}$ the controllable gate
voltage. When the Josephson coupling energy $E_{J}$ is much smaller than the
charging energy $E_{ch}$, and both of them are much smaller than the
superconducting energy gap $\Delta $, the Hamiltonian Eq.(\ref{e28}) can be
parameterized by the number of the Cooper pairs $n$ on the island as \cite%
{PRL1997} \cite{RMP2001}%
\begin{eqnarray}
H_{R} &=&\sum_{n}\left\{ E_{ch}\left( n-n_{g}\right) ^{2}\left\vert
n\right\rangle \left\langle n\right\vert \right.  \nonumber \\
&&\left. -\frac{1}{2}E_{J}\left[ \left\vert n\right\rangle \left\langle
n+1\right\vert +\left\vert n+1\right\rangle \left\langle n\right\vert \right]
\right\} .  \label{e29}
\end{eqnarray}%
When the temperature $T$ is low enough the system can be reduced to a
two-state system (qubit) because all other charge states have much higher
energy and can be neglected. So the Hamiltonian of the system can \emph{%
approximately} reads%
\begin{equation}
H_{s}=-\frac{1}{2}B_{z}\sigma _{z}-\frac{1}{2}B_{x}\sigma _{x}.  \label{e30}
\end{equation}%
Here, $B_{z}=E_{ch}\left( 1-2n_{g}\right) $ and $B_{x}=E_{J}.$ Eq.(\ref{e30}%
) is similar to the ideal single qubit model \cite{RMP2001}, but it can be
modulated only by changing one parameter $B_{z}.$ However, changing the
parameter $B_{z}$ (through switching the gate voltage) one can perform the
desirable one-bit operations. If, for example, one chooses the idle state
far to the left from the degeneracy point, the eigenstate lose to $%
\left\vert 0\right\rangle $ and $\left\vert 1\right\rangle .$ Then switching
the system suddenly to the degeneracy point for a time $t$ and back produces
a rotation in spin space \cite{RMP2001},%
\begin{equation}
U_{J}=\exp \left( \frac{iE_{J}t}{2}\sigma _{x}\right) =\left( 
\begin{tabular}{ll}
cos$\frac{tE_{J}}{2}$ & $i$sin$\frac{tE_{J}}{2}$ \\ 
$i$sin$\frac{tE_{J}}{2}$ & cos$\frac{tE_{J}}{2}$%
\end{tabular}%
\right) .  \label{e31}
\end{equation}%
This can be obtained by modulating the gate voltage and making $B_{z}=E_{ch}$
$\left( 1-2n_{g}\right) =0.$ Thus, $H_{S}=-\frac{B_{x}}{2}\sigma _{x},$ and
Eq.(\ref{e7}) becomes 
\begin{equation}
H=-\frac{B_{x}}{2}\sigma _{x}+\sigma _{z}\sum_{j}g_{j}^{\prime
}b_{j}^{\dagger }b_{j}+\sum_{j}\omega _{j}^{\prime }b_{j}^{\dagger }b_{j}.
\label{e32}
\end{equation}%
In the following, we will calculate the $\left\Vert \sigma \left( t\right)
\right\Vert _{\lambda }$ and the decoherence $D\left( t\right) $ of JCQ in
the system (\ref{e32}). In the calculation, three pure initial states are
chosen, they are $\left\vert \phi \right\rangle _{0}=\left( 1,0\right) ^{T},$
(corresponding to points in the following Figs.); $\left\vert \phi
\right\rangle _{1}=\left( \frac{\sqrt{3}}{2},\frac{1}{2}\right) ^{T},$
(corresponding to above lines); $\left\vert \phi \right\rangle _{2}=\left( 
\sqrt{\frac{1}{2}},\sqrt{\frac{1}{2}}\right) ^{T},$ (corresponding to below
lines), [because it has been shown that evaluation of the supremum over the
initial density operators in order to find $D\left( t\right) ,$ see Eq.(\ref%
{e27}) one can do over only pure-state density operators \cite{Privman}].
Here, the eigenstates and eigenvalues of $H_{s}$ are%
\begin{eqnarray}
\left\vert \varphi _{0}\right\rangle &=&\frac{1}{\sqrt{2}}\left( \left\vert
0\right\rangle -\left\vert 1\right\rangle \right) ,\text{ }\lambda _{0}=%
\frac{B_{x}}{2},  \nonumber \\
\left\vert \varphi _{1}\right\rangle &=&\frac{1}{\sqrt{2}}\left( \left\vert
0\right\rangle +\left\vert 1\right\rangle \right) ,\text{ }\lambda _{1}=-%
\frac{B_{x}}{2}.  \label{e33}
\end{eqnarray}%
The eigenstates and eigenvalues of $\sigma _{z}$ are%
\begin{eqnarray}
\left\vert \psi _{0}\right\rangle &=&\left\vert 0\right\rangle ,\text{ \ }%
\chi _{0}=1,  \nonumber \\
\left\vert \psi _{1}\right\rangle &=&\left\vert 1\right\rangle ,\text{ \ }%
\chi _{0}=-1.  \label{e34}
\end{eqnarray}%
Thus, by using Eq.(\ref{e23}) we can obtain%
\begin{equation}
\sigma _{11}=\frac{1}{4}\left( \rho _{00}-\rho _{11}\right) \left(
2-W_{1}-W_{2}\right) +\frac{1}{2}\rho _{10}\sin \frac{tE_{J}}{2}\left(
W_{2}-W_{1}\right) ,  \label{e35}
\end{equation}%
\begin{eqnarray}
\sigma _{10} &=&\frac{1}{4}\left( \rho _{00}-\rho _{11}\right) e^{\frac{%
itE_{J}}{2}}\left( W_{1}-W_{2}\right)  \nonumber \\
&&+\frac{1}{4}\rho _{10}\left[ 2\left( 1+e^{itE_{J}}\right) +\left(
W_{1}+W_{2}\right) \left( e^{itE_{J}}-1\right) \right] ,  \label{e36}
\end{eqnarray}%
where 
\begin{equation}
W_{1}=\dprod\limits_{j=N_{0}}^{N}\frac{-1+e^{-\beta \omega _{j}}}{%
-1+e^{i2g_{j}^{\prime }t-\beta \omega _{j}}},W_{2}=\dprod%
\limits_{j=N_{0}}^{N}\frac{-1+e^{-\beta \omega _{j}}}{-1+e^{-i2g_{j}^{\prime
}t-\beta \omega _{j}}}.  \label{e37}
\end{equation}%
In the following, we numerically investigate the decoherence of the JCQ in
this model. We choose $E_{J}=51.8\mu ev$ according to \cite{Nakamura}, and $%
T=30mK.$ Suppose the frequencies of the harmonic oscillators made up of bath
are range from $\omega _{low}=1$ $MHz$ to $\omega _{high}=20$ $MHz\equiv
\omega _{c}$ (corresponding $N_{0}=1,$ $N=20$). Thus, we can plot the norms $%
\left\Vert \sigma \right\Vert _{\lambda }$ versus time $t$ in Fig.1. It is
shown that when the initial state is $\rho \left( 0\right) =\left\vert \phi
\right\rangle _{00}\left\langle \phi \right\vert ,$ $\left\Vert \sigma
\left( t\right) \right\Vert _{\lambda }$ becomes maximum and it equals to $%
D\left( t\right) $ (plotted by points in the Figs.). We denote the low
decoherence ($D\left( t\right) \leq 10^{-4}$) time $t^{ld}.$ From Fig.1b we
obtain $t_{1}^{ld}\approx 8\times 10^{-5}\times 6.582\times
10^{-10}s=5.266\times 10^{-2}ps.$ We denote the elementary gate operation
time, the characteristic time $t^{g}.$ In this case, $t_{1}^{g}=\hbar
/E_{J}=1.27\times 10^{1}ps$. Thus, $t_{1}^{ld}\ll t_{1}^{g}.$ It shows that
at the temperature $T=30mK$ the present setup of the JCQ cannot be taken as
the qubit for quantum computation because it do not satisfy the DiVincenzo
low decoherence criterion \cite{DiVincenzo}. However, if we make the
temperature $T$ be lowered, the decoherence will decreases. For example,
setting $T=0.3mK$, and keeping $E_{J}=51.8\mu ev,$ we can plot the norms $%
\left\Vert \sigma \right\Vert _{\lambda }$ versus time $t$ in Fig.2. In this
case we can obtain $t_{2}^{ld}\approx 2\times 10^{-2}\times 6.582\times
10^{-10}s=13.16ps$ and $t_{2}^{g}=t_{1}^{g}=12.7ps,$ where $%
t_{2}^{ld}>t_{2}^{g}$. It is shown that when the temperature decreases to $%
T=0.3mK$, within the whole time of elementary gate operation, $D\left(
t\right) \leq 10^{-4}.$ Theoretically, in the lower temperature the JCQ in
nonlinearly coupling JCQ-bath becomes an ideal qubit for making quantum
computer because it satisfy the DiVincenzo low decoherence criterion. 
\begin{eqnarray*}
&& \\
&& \\
&&Fig.1a,\text{ }Fig.1b \\
&& \\
&& \\
&&
\end{eqnarray*}%
\emph{Fig.1: Norms }$\left\Vert \sigma \right\Vert _{\lambda }$\emph{\
versus time }$t,$\emph{\ (a) in a longer time comparing to the elementary
gate operation time; (b) in the low decoherence (}$D\left( t\right) \leq
10^{-4}$\emph{) time. Here, the points and lines correspond to different
initial states (see the explanation in text), }$E_{J}=51.8\mu ev$\emph{, }$%
T=30mk$\emph{\ and }$N_{0}=1,$\emph{\ }$N=20.$\emph{\ The unit of the time
is }$6.582\times 10^{-10}s.$%
\begin{eqnarray*}
&& \\
&& \\
&& \\
&&Fig.2a,\text{ }Fig.2b \\
&& \\
&&
\end{eqnarray*}%
\emph{Fig.2: Norms }$\left\Vert \sigma \right\Vert _{\lambda }$\emph{\
versus time }$t,$\emph{\ (a) in a longer time comparing to the elementary
gate operation time; (b) in the low decoherence (}$D\left( t\right) \leq
10^{-4}$\emph{) time.\ Here, the values and meanings of the parameters are
same as Fig.1 expect for }$T=0.3mk.$

\section{Conclusions}

In this Letter we firstly constructed a qubit-bath model where we supposed
that the qubit nonlinearly couples with its environment. Then using the
model we investigated the short-time decoherence of the JCQ. The decoherence
is described by the norm of the deviation density operator, $\left\Vert
\sigma \left( t\right) \right\Vert _{\lambda }.$ We define the biggest $%
\left\Vert \sigma \left( t\right) \right\Vert _{\lambda }$ the quantity of
decoherence $D\left( t\right) $. In the previous paper \cite{Liang01} by
using the same setup of the JCQ and bath's parameters we calculated the
decoherence of the JCQ, where the JCQ is supposed linearly coupling with its
bath. In the linearly coupling model, the decoherence is not larger than the
DiVincenzo low decoherence criterion so the JCQ can be taken as the qubit
for quantum computations. However, the results of this Letter show that at
the same experimental temperature $T=30mk,$ when the JCQ nonlinearly couples
with its bath, it is not a ideal block to build the quantum computer because
the decoherence is larger than the DiVincenzo low decoherence criterion. But
when the temperature decreases to $T=0.3mk$ the decoherence decreases to a
endurable grade. It suggests that in order to make the JCQ become a block of
quantum computing hardware developing the technique of low temperature may
be important. It should be noted that this is a pilot study to the
decoherent problem of JCQ. The frequencies of the harmonic oscillators of
the bath modes are chosen tentatively. They may only be obtained by virtue
of some ingenious experiments. A further numerical simulation shows that
when the frequencies are larger than our setting the JCQ will adapt a higher
temperature. But, no matter what they are, we can find out a proper
temperature for the JCQ in low decoherence by using our method.

Acknowledgement: \emph{This work was supported by the National Natural
Science Foundation of China (NSFC), grant Nos. 10347133, 10347134 and Ningbo
Youth Foundation, grant Nos. 2004A620003 and 2003A620005. X. T. L thanks
Jiang Wei (Univ. of Washington) for his help.}

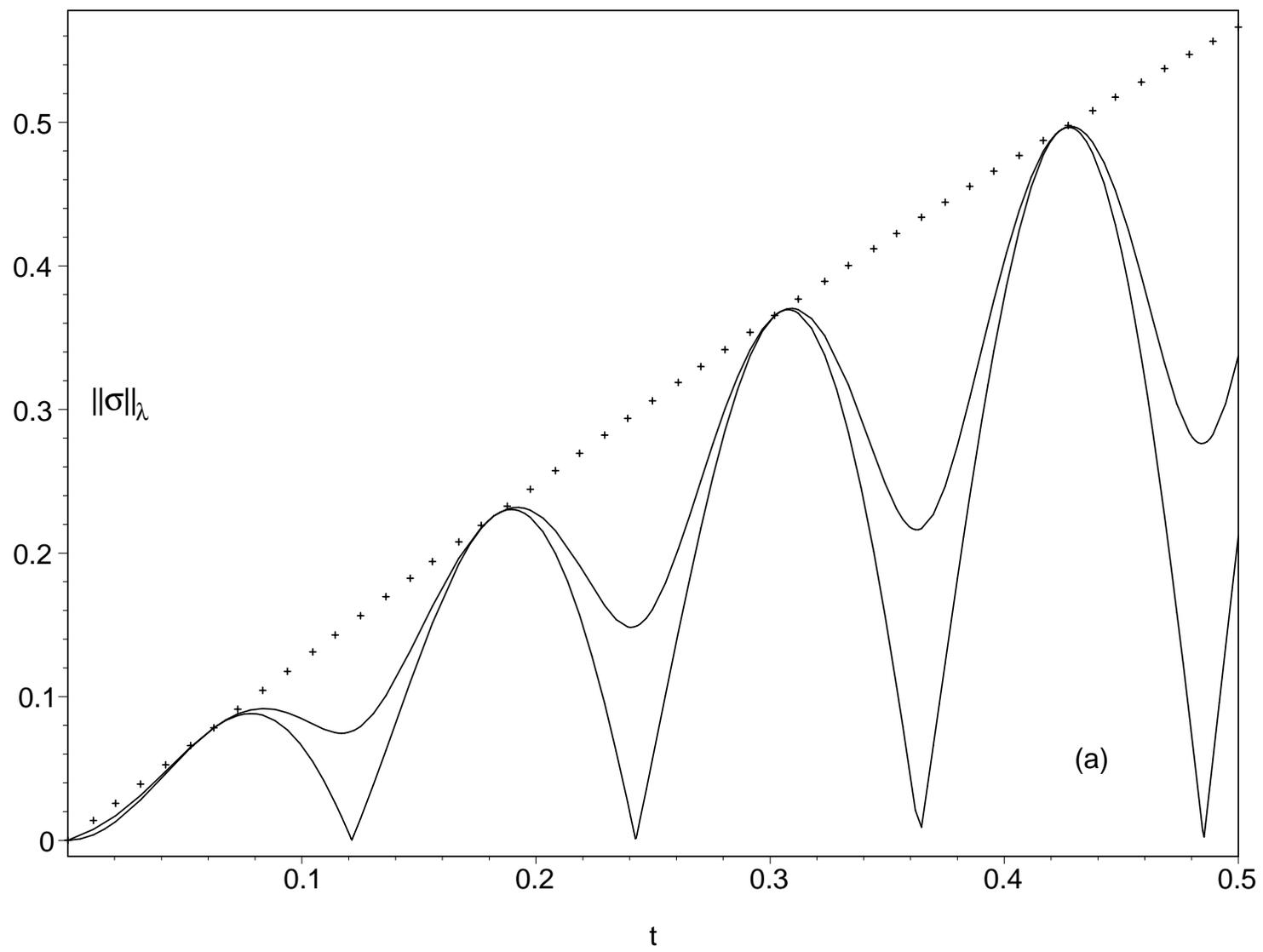
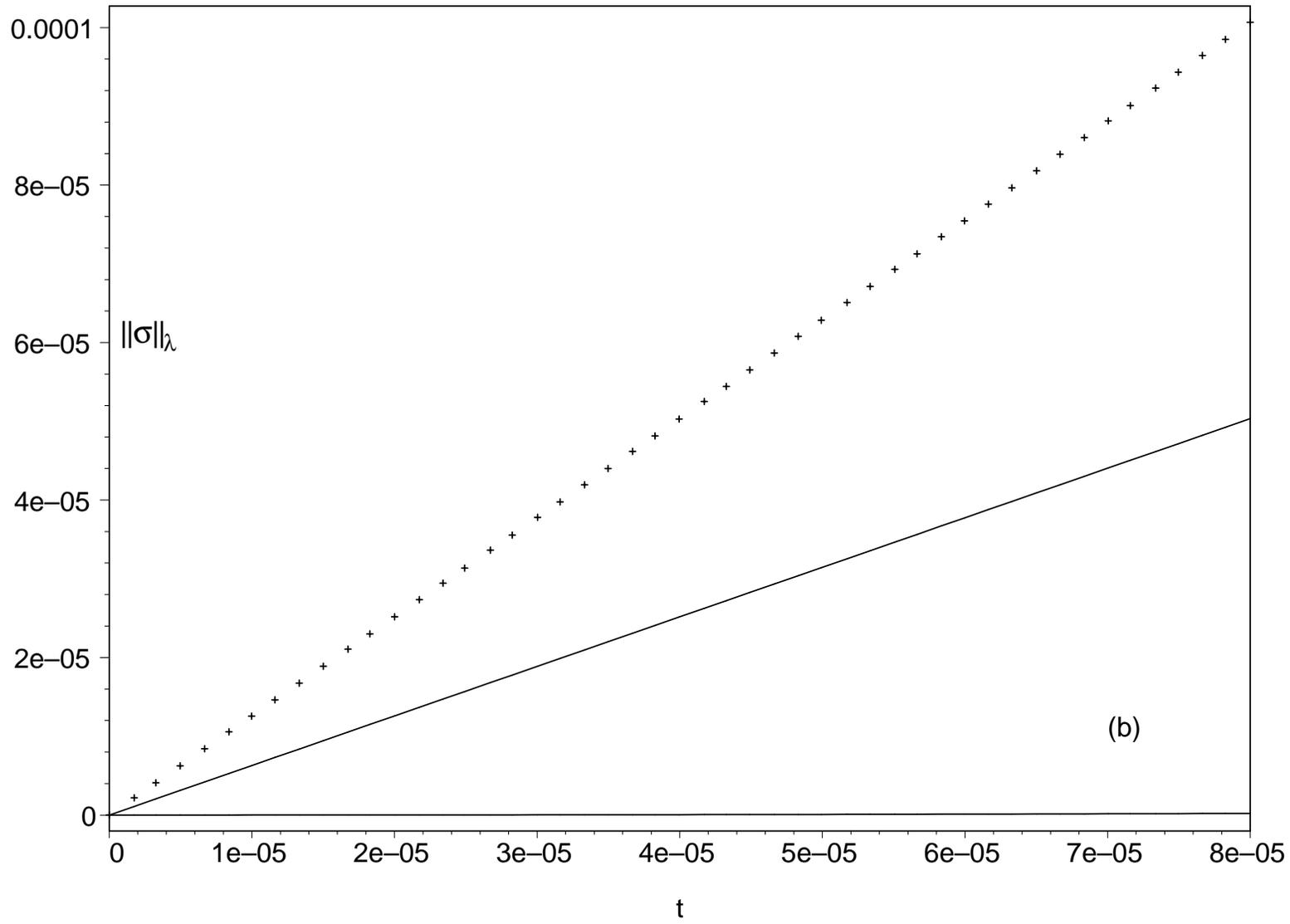
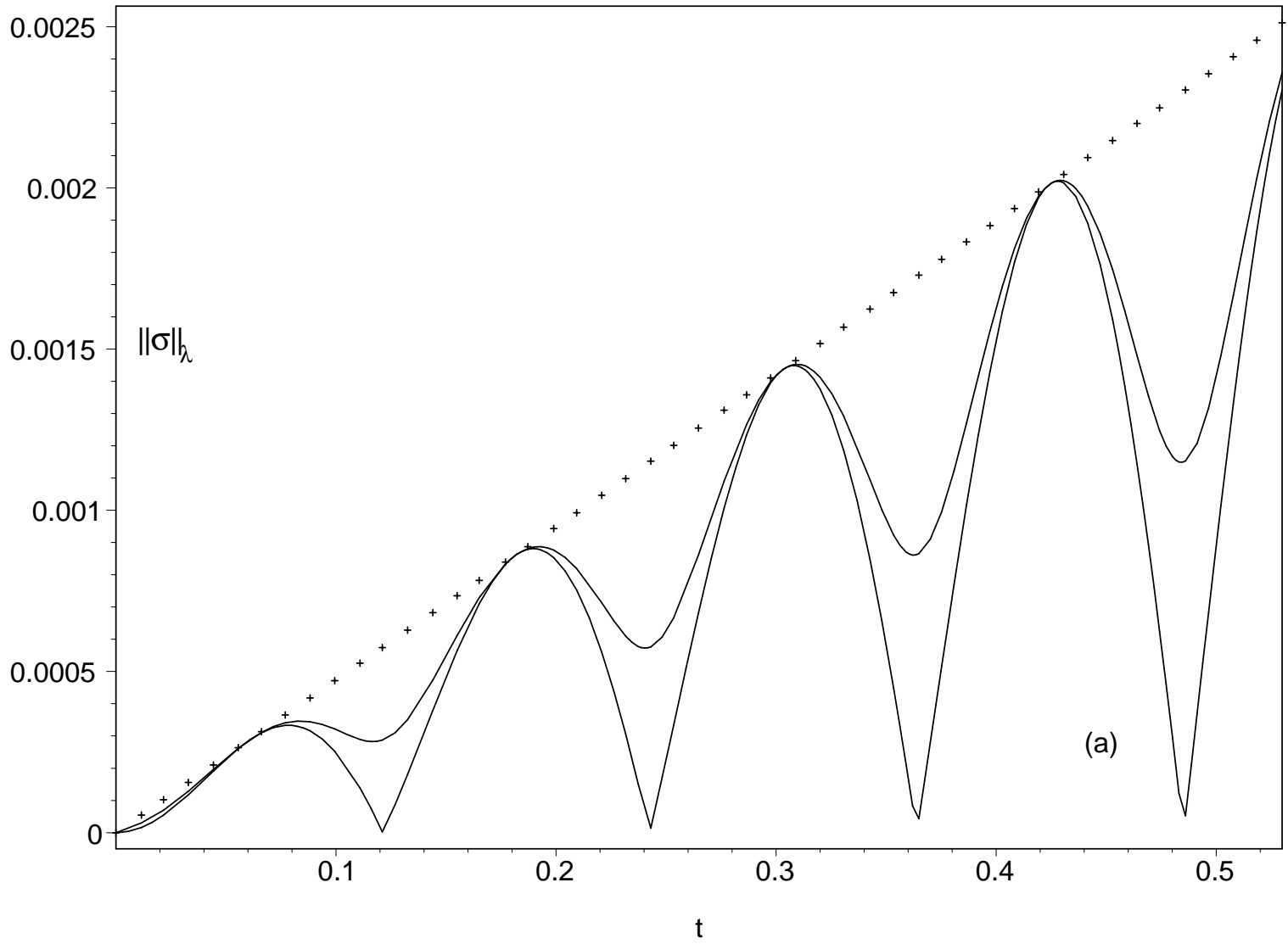
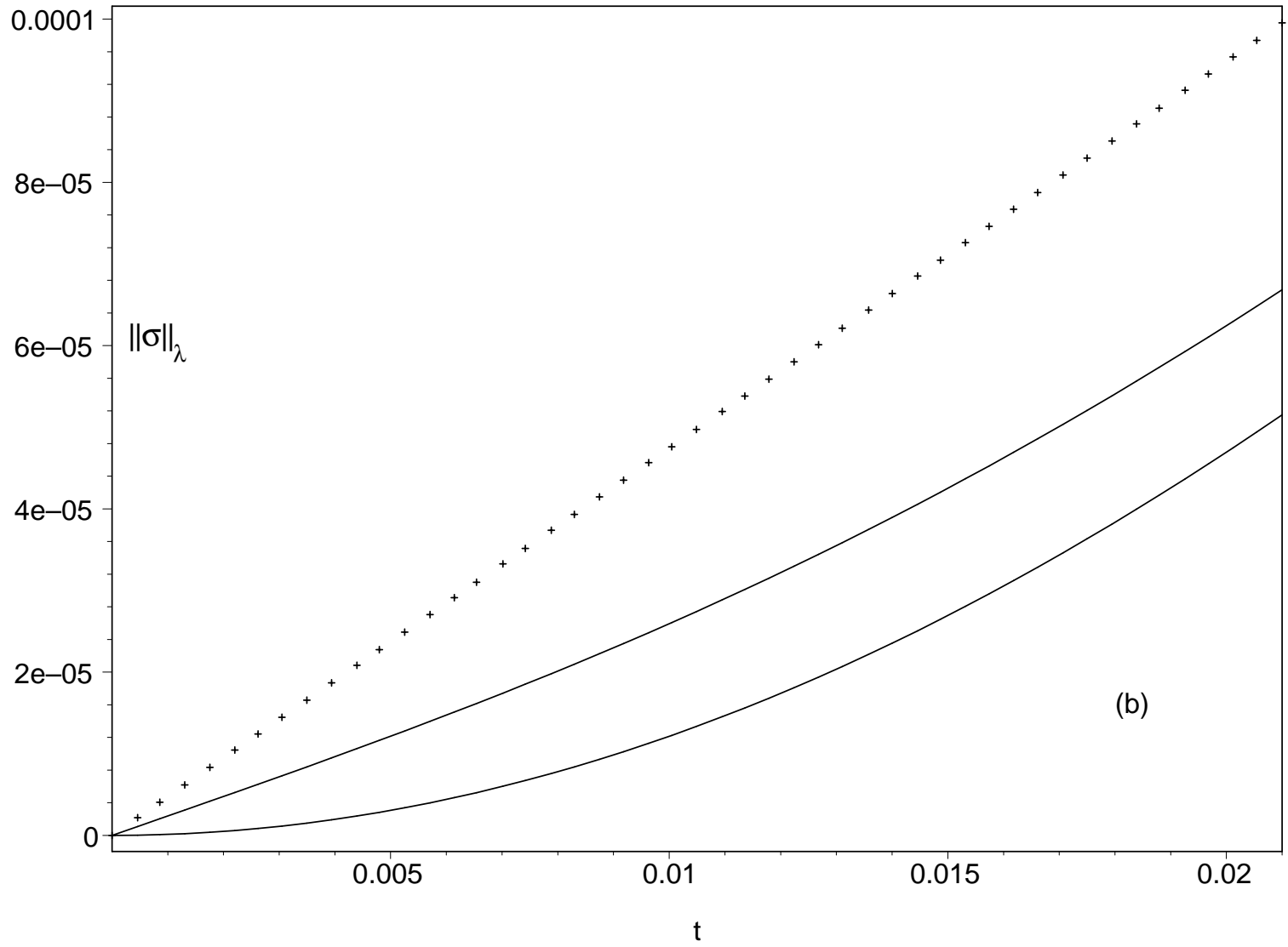
\end{document}